\documentclass[%
 reprint,
 amsmath,amssymb,
 aps,
dvipdfmx
]{revtex4-1}
\usepackage{graphicx}
\usepackage{dcolumn}
\usepackage{bm}

\usepackage{color}
\newcommand{\red}[1]{{#1}}
\renewcommand{\a}{\alpha}

\newcommand{\bea}{\begin{eqnarray}}
\newcommand{\eea}{\end{eqnarray}}
\newcommand{\f}[2]{\frac{#1}{#2}}
\newcommand{\eq}{&=&}
\newcommand{\nn}{\nonumber \\ }
\newcommand{\ve}{\varepsilon}
\renewcommand{\d}{\delta}
\newcommand{\area}{\int_{-\infty}^\infty }

\renewcommand{\l}{\lambda}
\newcommand{\p}{\partial}
\newcommand{\pp}[2]{\f{\p #1}{\p #2}}

\newcommand{\sref}[1]{{Eq.} (\ref{#1})}
\newcommand{\pmi}{+i\ve}
\newcommand{\citeauthorname}[2]{{#1} {#2}}
\newcommand{\citebook}[4]{{#1} {\it #2} ({#3}, {#4}).}
\newcommand{\citepaper}[4]{{#1} {#3} ({#4}).}
\begin{document}

\preprint{APS/123-QED}

\title{
Asymptotic Eigenvalue Distribution of Wishart Matrices whose Components 
are not Independently and Identically Distributed}

\author{Takashi Shinzato}
\email{takashi.shinzato@r.hit-u.ac.jp}
 \affiliation{
Mori Arinori Center for Higher Education and Global Mobility,
Hitotsubashi University, 
Tokyo, 1868601, Japan.}

\date{\today}

\begin{abstract}
In the present work, eigenvalue distributions defined by a random rectangular matrix whose components are 
neither independently nor identically distributed are analyzed using 
 replica analysis and belief propagation. In particular, we consider the 
 case in which the components are independently but not identically 
 distributed; for example, only the components in each row or in each 
 column may be {identically distributed}. We also consider the more 
 general case in which the components are correlated with one another. 
 We use the replica approach \red{while making only weak assumptions} in order to determine the asymptotic eigenvalue distribution and to derive an algorithm for doing so, based on belief propagation. One of our findings supports the results obtained from Feynman diagrams. We present the results of several numerical experiments that validate our proposed methods.
\begin{description}
\item[PACS number(s)]
{89.90.+n}, 
{75.10.Nr},
{89.65.Gh}, 
{02.50.-r}

\end{description}
\end{abstract}
\maketitle


\section{Introduction}
Random matrices, in which each component is regarded as a random variable, are widely 
used and investigated, both theoretically and practically, in many fields of research, including number theory, combinatorial theory, nuclear physics, 
condensed matter physics, bionomics, mathematical finance, and 
communication theory \red{\cite{Mehta,Tulino,Bai,Guhr,Kwapien1}}. In particular, 
the mathematical structure of random square matrices has been investigated; 
topics of investigation include the 
eigenvalue distribution and distribution of the level spacings of a 
Gaussian unitary ensemble (GUE) characterized by a Hermitian random matrix, and those of a 
Gaussian orthogonal ensemble (GOE) characterized by an orthogonal random
matrix. 
For random rectangular matrices, topics of investigation have included 
singular values and the asymptotic 
eigenvalue distribution of a Wishart matrix that is defined by an 
autocovariance matrix 
\cite{MP,Silverstein1,Silverstein2,Sengupta,Burda1,Burda2,Recher}. 
For instance, 
Mar$\check{\rm c}$enko and Pastur consider the asymptotic eigenvalue distribution when each entry of a given random rectangular matrix is independently 
and identically drawn from a population with
a probability distribution with mean $0$ and variance $1/N$, 
and  $N \times N$ { autocovariance matrix}, and the eigenvalue distribution of the autocovariance matrix 
is sufficiently close to the asymptotic distribution when it is sufficiently large; this is known as the Mar$\check{\rm c}$enko--Pastur law \cite{MP}. 
Silverstein and Choi used the Stieltjes transformation to rederive the asymptotic eigenvalue distribution for the 
Mar$\check{\rm c}$enko--Pastur law 
\cite{Silverstein1,Silverstein2}. 
Sengupta and Mitra  
expanded the resolvent of a random matrix in which the components are correlated with 
one another, using the inverse of the matrix size, 
and they used Feynman diagrams to derive the fixed-point equations 
that would determine the asymptotic 
eigenvalue distribution \cite{Sengupta}.
Burda, G$\ddot{\rm o}$rlich, Jarsoz, and Jukiewicz
derived the relationship between 
the asymptotic eigenvalue distribution of a correlation matrix, 
that was obtained using Feynman diagrams and
an eigenvalue distribution estimated from a practical dataset \cite{Burda1}.
In addition, 
Burda, Jurkiewicz, and Waclaw
successfully derived 
the relation between those estimates by considering the resolvent and a
moment-generating function \cite{Burda2}. 
Recher, Kieburg, and Guhr used supermatrix theory to assess the eigenvalue distribution of small, random 
matrices in which the components were correlated, 
and they compared the theoretical results with those obtained from numerical 
experiments 
\cite{Recher}.

As discussed above, there have been many studies that use Feynman diagrams 
or supermatrix theory to evaluate the asymptotic eigenvalue distribution defined 
by a random matrix ensemble, but 
few studies have used replica analysis or belief propagation to
 investigate the 
asymptotic eigenvalue distribution of a Wishart matrix in which the
components are independently but not identically distributed, or in which they are correlated with each other.  
It has been assumed that the resolvent can be expanded to the inverse of the matrix size and that the ensemble average of each term is independent; in addition, it has been implicitly assumed that, in the Feynman diagram approach, a 
recursive relation with respect to the
irreducible self-energy is 
a primary part of the resolvent.
Moreover, 
since it is necessary to compute the inverse matrix in order to use the Feynman 
diagram approach, the required computational time is excessive.

{We note }that the 
portfolio optimization problem is widely considered to be one of the most important applications of random matrix theory. If we consider an investment market in which the variance of the return rate of 
assets is not identical, 
then we need to use 
a random matrix ensemble in which the 
components are not identically distributed \red{\cite{Markowitz,Wakai,Shinzato-SA2015,Shinzato-Yasuda-BP2015,Shinzato-heteroMV2016,Kwapien,Drozdz}}. Furthermore, 
since we can use the asymptotic eigenvalue distribution defined by the Wishart matrix to evaluate the typical behavior of two quantities that characterize the optimal portfolio (defined as the portfolio that  
minimizes the investment risk), in order to solve the portfolio optimization problem, we need to systematically examine the asymptotic eigenvalue distribution of nonidentically distributed 
random matrices. 
Thus, in the present paper, 
our goal is to determine the asymptotic eigenvalue distribution of a random matrix ensemble in which the entries are independently but not 
identically distributed or in which they are correlated with one another; we will do this using 
replica analysis, since it does {not} require the computation of the inverse matrix, 
and belief propagation, which does not require second-order 
statistics. We verify the effectiveness of 
our proposed method by presenting the results of several numerical experiments.

This paper is organized as follows. In Section II,  
we consider the relationship between Green's function and the eigenvalue 
distribution; we do this in order to analytically derive the asymptotic
 eigenvalue distribution and to explain the approach used in various previous studies.
In Section \ref{sec3}, we develop a methodology based on 
replica analysis in order to evaluate the asymptotic eigenvalue distribution 
of a random matrix ensemble in which the entries are {neither 
independently nor identically distributed}. In 
Section \ref{sec4}, in a similar way, we derive an algorithm based on belief propagation. In Section \ref{sec5}, we present the results of
 numerical simulations that show the 
consistency and accuracy of our proposed methods. {Section VI is 
devoted to a summary of our findings and a discussion of areas for 
future work}.


\section{Eigenvalue distributions and Green's functions\label{sec2}}
\subsection{Asymptotic eigenvalue distribution and Green's functions}
In this subsection, 
as a preparation for deriving the asymptotic eigenvalue distribution, 
we discuss the relationship between the eigenvalue distribution of a Wishart 
matrix defined by a random rectangular matrix and Green's functions. 
Similar to the discussion of Wishart et al. {\cite{Mehta,Wishart,Fisher}}, {we consider a random rectangular matrix, $X=\left\{\f{x_{i\mu}}{\sqrt{N}}\right\}\in{\bf R}^{N\times p}, 
(i=1,\cdots,N,\mu=1,\cdots,p)$.} For simplicity, we will assume that by random matrix we mean a 
random rectangular matrix; 
we will assume that the expectation of each entry of a random 
matrix is (or is normalized to be) $0$; and we will assume that $1/\sqrt{N}$ is 
a scaling coefficient determined by the maximum or minimum eigenvalue 
of the variance-covariance matrix, which is a random matrix (Wishart matrix) in which 
$\a=p/N\sim O(1)$. From 
these settings, the eigenvalue distribution of the Wishart matrix $XX^{\rm 
T}\in{\bf R}^{N\times N}$, $\rho(\l|X)$, 
can be written using the $N$ eigenvalues $\l_1,\cdots,\l_N$ as follows:
\bea
\rho(\l|X)\eq\f{1}{N}\sum_{k=1}^N\d(\l-\l_k),
\label{eq1}
\eea
where $\d(x)$ is the Dirac delta function, {and a superscript ${\rm T}$ indicates 
the transposition of a vector or matrix.} In addition, by using the trace operator, ${\rm 
Tr}$, \sref{eq1} can be rewritten as $\rho(\l|X)=
\f{1}{N}{\rm Tr}\d\left(\l I_N-XX^{\rm T}\right)$, 
where $\d(Y)=
\lim_{\ve\to+0}\f{1}{2\pi i}
\left((Y-i\ve I_N)^{-1}-(Y+i\ve I_N)^{-1}
\right)$, $Y\in{\bf R}^{N\times N}$;  and $I_N$ is the identity matrix, i.e., $I_N={\rm diag}\{1,1,\cdots,1\}\in{\bf R}^{N\times N}$ (hereafter, $I_m\in{\bf R}^{m\times m}$ 
will be used to denote the $m$-dimensional identity matrix).

Next, in order to derive the eigenvalue distribution, 
we define two kinds of Green's function (or resolvent), as follows:
\bea
\label{eq5}
G^R(\l|X)
\eq
\lim_{\ve\to+0}
\f{1}{N}
{\rm Tr}\left((\l+i\ve)I_N-XX^{\rm T}\right)^{-1},\\
\label{eq4}
G^A(\l|X)
\eq
\lim_{\ve\to+0}
\f{1}{N}
{\rm Tr}\left((\l-i\ve)I_N-XX^{\rm T}\right)^{-1},
\eea
where 
$G^R(\l|X)$ is the retarded Green's function, and 
$G^A(\l|X)$ is the advanced Green's function. 
From these definitions, 
we can have the following relations for the real and imaginary parts of these Green's functions: 
${\rm Re}G^{R}(\l|X)={\rm Re}G^{A}(\l|X)$, and $
{\rm Im}G^{R}(\l|X)=-{\rm Im}G^{A}(\l|X)$ 
From this, the eigenvalue distribution of a Wishart matrix 
$XX^{\rm T}\in{\bf R}^{N\times N}$, $\rho(\l|X)$, can be rewritten using $G^R(\l|X)$ and $G^A(\l|X)$ as follows:
\bea
\rho(\l|X)\eq
-
\f{1}{2\pi i}\left(G^{R}(\l|X)-G^{A}(\l|X)\right)\nn
\eq-\f{1}{\pi}{\rm Im}G^{R}(\l|X).\label{eq7}
\eea
From \sref{eq7}, it can be seen that 
if we could analytically assess the retarded Green's function $G^R(\l|X)$, then 
we could derive the eigenvalue distribution $\rho(\l|X)$ from its imaginary part.

Finally, when $N$ is sufficiently 
large, the asymptotic eigenvalue distribution 
$\rho(\l)$ is {said to be} self-averaging, that is, $\rho(\l|X)=E_X[\rho(\l|X)]$, {
$E_X[f(X)]$ means the expectation of $f(X)$ on random variables $X$.}
Thus, we will not analyze  
the eigenvalue distribution of a Wishart matrix $\rho(\l|X)$, 
but instead, we will determine its asymptotic eigenvalue distribution, $\rho(\l)=E_X[\rho(\l|X)]$.
\subsection{Previous studies}
We now present some findings obtained in previous studies for the 
asymptotic eigenvalue distribution. Several previous studies have considered 
the case in which each entry, $x_{i\mu}$, of a random matrix is 
{independently and identically distributed.} For example, when the distribution of the entries has mean $0$ and 
variance $1$, and $\a=p/N$, 
the asymptotic eigenvalue distribution (for large $N$) 
converges to the
Mar$\check{\rm c}$enko-Pastur law, as follows:
\bea
\rho(\l)\eq[1-\a]^+\d(\l)+\f{\sqrt{[\l_+-\l]^+[\l-\l_-]^+}}{2\pi\l},
\eea
where $\l_\pm=(1\pm\sqrt{\a})^2$, and $[u]^+=\max(u,0)$ \cite{MP}.

In a more general case, 
which will be discussed in detail below, 
if the asymptotic eigenvalue distribution of a random 
matrix ensemble has 
$E_X[x_{i\mu}x_{j\nu}]=m_{ij}\theta_{\mu\nu}$, 
the asymptotic eigenvalue distribution can be derived
by expanding the generating function in terms of the characteristic parameters.
From this property, previous studies have
determined the asymptotic eigenvalue distribution 
by using recursive relations with respect to the $1/N$ expansion of the generating 
functions based on Feynman diagrams \cite{Burda1}.
There are two key properties: (1) Green's functions are related to self-energy, and (2)
self-energy can be divided into irreducible self-energy; from these properties, we can obtain
simultaneous equations for the order parameters, as follows \cite{Sengupta}:
\bea
\label{eq10}
\label{eq12}
Q_w\eq\left((\l\pmi) I_N+M\left(\f{1}{N}{\rm Tr}Q_t\right)\right)^{-1},\\
\label{eq13}
Q_s\eq MQ_w,\\
\label{eq14}
Q_u\eq\left(\Theta\left(\f{1}{N}{\rm Tr}Q_s\right)
-I_p\right)^{-1},\\
\label{eq15}
Q_t\eq \Theta Q_u,
\eea
where 
$E_X[x_{i\mu}x_{j\nu}]=m_{ij}\theta_{\mu\nu}$, 
$M=\left\{m_{ij}\right\}\in{\bf R}^{N\times N}$, and 
$\Theta=\left\{\theta_{\mu\nu}\right\}\in{\bf R}^{p\times p}$ are 
replaced by the 
$N$-dimensional matrices $Q_w,Q_s$ and  
$p$-dimensional matrices $Q_u,Q_t$. We can solve these 
simultaneous equations to obtain the asymptotic eigenvalue distribution:
\bea
\rho(\l)\eq-\f{1}{\pi}{\rm Im}\lim_{\ve\to+0}\f{1}{N}{\rm Tr}Q_w.
\eea
Comparing this with the Feynman diagram method, we note that
$Q_w,Q_u$ correspond to Green's functions  
and 
$Q_s,Q_t$ correspond to irreducible self-energy.

In order to simplify the calculation of \sref{eq12} to \sref{eq15}, 
we rewrite them as follows:
\bea
\chi_w\eq\f{1}{N}{\rm Tr}Q_w,\\
\chi_s\eq\f{1}{N}{\rm Tr}Q_s,\\
\chi_u\eq\f{1}{p}{\rm Tr}Q_u,\\
\chi_t\eq\f{1}{p}{\rm Tr}Q_t.
\label{eq18}
\eea
Using the novel order parameters $\chi_w,\chi_s,\chi_u,\chi_t\in{\bf C}$, we can write the following
simultaneous equations: 
\bea
\label{eq21}\chi_w\eq\f{1}{N}
{\rm Tr}\left((\l\pmi) I_N+\a\chi_t M\right)^{-1},\\
\label{eq20}
\chi_s
\eq
\f{1-(\l\pmi)\chi_w}{\a\chi_t},\\
\label{eq21-2}
\chi_u\eq\f{1}{p}
{\rm Tr}
\left(\chi_s\Theta-I_p\right)^{-1},\\
\chi_t
\eq\f{1+\chi_u}{\chi_s}.\label{eq24}
\label{eq22}
\eea
Then, from $\chi_w$, we have
\bea
\rho(\l)\eq-\f{1}{\pi}{\rm Im}\lim_{\ve\to+0}\chi_w,
\eea
that is, the limit of one of these parameters gives the asymptotic eigenvalue 
distribution. These newly defined parameters allow us to solve the simultaneous equations comparatively easily, compared to solving the original matrix formulation.
However, it is still necessary to calculate the inverse matrix  
in \sref{eq21} and \sref{eq21-2}, and thus 
it is difficult to implement this approach 
{and to calculate the inverse matrices in \sref{eq10} to 
\sref{eq15} when $N,p$ are large \cite{Sengupta,Burda1}.}
It is not sufficient to discuss the adequacy of 
the assumption 
that we can expand $Q_w,Q_s,Q_u,Q_t$ over $1/N$, based on the Feynman diagram 
method. Thus, 
we will use replica analysis for a quenched ordered system in order to directly solve the asymptotic eigenvalue distribution of a
random matrix ensemble; this approach will not require the 
calculation of an inverse matrix, and it validates the adequacy of 
their {approaches \cite{Sengupta,Burda1}}. As an alternative
approach, 
we propose an algorithm based on the belief propagation method; this approach allows us to determine 
the eigenvalue distribution 
without the need to calculate an inverse matrix,  
when $N,p$ are sufficiently large but not infinite.

\section{Replica analysis\label{sec3}}
\subsection{Replica trick}
We now discuss the use of replica analysis to solve the asymptotic eigenvalue 
distribution $\rho(\l)$; this is done in a way to similar to that presented in previous studies \cite{Kamenev,Edwards,Dhesi,Nishimori}.
We can rewrite the retarded Green's function as 
\bea
G^R(\l|X)\eq-2\lim_{\ve\to+0}\pp{\phi(\l\pmi|X)}{\l},
\label{eq24-22}
\eea
where the partition function $Z(\l\pmi|X)$ and the generating 
function $\phi(\l\pmi|X)$ are defined as follows:
\bea
\label{eq26}
Z(\l\pmi|X)\eq\det\left|
(\l\pmi)I_N
-XX^{\rm T}\right|^{-\f{1}{2}},
\\
\phi(\l\pmi|X)\eq
\f{1}{N}
\log Z(\l\pmi|X).
\eea
From \sref{eq7} and \sref{eq24-22}, the 
eigenvalue distribution can be derived as 
\bea
\label{eq27}
\rho(\l|X)
\eq\f{2}{\pi}{\rm Im}\lim_{\ve\to+0}\pp{\phi(\l\pmi|X)}{\l}.
\eea
Moreover, since its asymptotic eigenvalue distribution $\rho(\l)$ is 
evaluated as 
\bea
\label{eq30}
\label{eq28}
\rho(\l)\eq E_X[\rho(\l|X)]\nn
\eq
\f{2}{\pi}{\rm Im}\lim_{\ve\to+0}
\pp{}{\l}E_X[\phi(\l\pmi|X)],
\eea
in order to implement \sref{eq30}, we need to 
assess 
\bea
E_X[\phi(\l\pmi|X)]\eq\f{1}{N}
E_X[\log Z(\l\pmi|X)].
\eea
That is, we need to average the generating function $\phi(\l\pmi|X)$
over all configurations of the random matrix $X$. 
We note that, in general, 
it is more difficult to 
assess the configurational average of the logarithm of the 
partition function. Thus, we can use an identity function known  
as the replica trick, $\log Z=\lim_{n\to0}\f{Z^n-1}{n}$, and we obtain
\bea
E_X[\log Z(\l\pmi|X)]\eq\lim_{n\to0}\f{E_X[Z^n(\l\pmi|X)]-1}{n};
\label{eq32}
\qquad
\eea
from this, we can compute the configurational average of the logarithm of the partition 
function  $E_X[\log Z(\l\pmi|X)]$
from the configurational average of a power function of the partition function
$E_X[Z^n(\l\pmi|X)]$. Moreover, using l'Hopital's rule with respect to \sref{eq32}, the replica 
trick can be rewritten as
\bea
\label{eq33}
E_X[\log Z(\l\pmi|X)]\eq\lim_{n\to0}\pp{}{n}\log E_X[Z^n(\l\pmi|X)],\nn
\eea
where, from the definition in \sref{eq26}, the 
partition function is 
\bea
Z(\l\pmi|X)
=\area \f{d\vec{w}}{(2\pi)^{\f{N}{2}}}
e^{
-\f{1}{2}
\vec{w}^{\rm T}
\left((\l\pmi)I_N
-XX^{\rm T}\right)
\vec{w}
}.\qquad
\eea
Furthermore, when the power $n$ is a natural number, 
one can expand the power function of the partition function in order to assess the configurational average comparatively easily:
\bea
&&
E_X\left[Z^n(\l\pmi|X)\right]\nn
\eq
E_X\left[
\area \prod_{a=1}^n
\f{d\vec{w}_a}{(2\pi)^{\f{Nn}{2}}}
e^{
-\f{1}{2}
\sum_{a=1}^n
\vec{w}_a^{\rm T}
\left((\l\pmi)I_N
-XX^{\rm T}\right)
\vec{w}_a
}
\right].
\nn
\eea

Thus, we can (comparatively) easily evaluate $E_X[Z^n(\l\pmi|X)]$ for $n\in{\bf N}$ 
with respect to each of the three statistical properties of each component 
of the random matrix, 
and thus, we can determine the asymptotic eigenvalue distribution.

\subsection{Independent but not identically distributed; case 1}
We consider the case in which each entry, $x_{i\mu}$, of the random matrix is distributed such that  
the probability has covariance $E_X[x_{i\mu}x_{j\nu}]= 
s_i\d_{ij}\d_{\mu\nu}$, and the higher-order moments are finite. That is, 
from $E_X[x_{i\mu}x_{j\nu}]= 
s_i\d_{ij}\d_{\mu\nu}$, we have 
\bea
\label{eq36}
M\eq{\rm diag}
\left\{s_1,s_2,\cdots,s_N\right\}\in{\bf R}^{N\times N},\\
\Theta\eq I_p\in{\bf R}^{p\times p}.\label{eq37}
\eea
We prepare the
order parameters: 
\bea
\label{eq38}
q_{wab}\eq\f{1}{N}\sum_{i=1}^Nw_{ia}w_{ib},
\\
q_{sab}\eq\f{1}{N}\sum_{i=1}^Nw_{ia}w_{ib} s_i,
\eea
and the conjugate order parameters: 
$\tilde{q}_{wab},\tilde{q}_{sab}$, $(a,b=1,2,\cdots,n)$. In this setting, 
using $n$-dimensional square matrices 
$Q_w,Q_s,\tilde{Q}_w,\tilde{Q}_s\in{\bf C}^{n\times n}$
whose components are the order parameters 
$q_{wab},q_{sab},\tilde{q}_{wab},\tilde{q}_{sab}$, we have
\bea
&&
\label{eq38-1}
\lim_{N\to\infty}\f{1}{N}\log E_X\left[Z^n(\l\pmi|X)\right]\nn
\eq\mathop{\rm 
Extr}_{Q_w,\tilde{Q}_w,Q_s,\tilde{Q}_s}\left\{
-\f{\a}{2}\log\det\left|I_n-Q_s\right|
+\f{1}{2}{\rm Tr}Q_w\tilde{Q}_w
\right.\nn
&&
\left.
+\f{1}{2}{\rm 
Tr}Q_s\tilde{Q}_s
-\f{1}{2}\left\langle
\log\det|
(\l\pmi)I_n+
\tilde{Q}_w+ s\tilde{Q}_s|
\right\rangle_{ s
}
\right\},\nn
\eea
where 
$\a=p/N\sim O(1)$ and 
\bea
\left\langle f( s)\right\rangle_{ s}\eq\lim_{N\to\infty}\f{1}{N}
\sum_{i=1}^Nf( s_i).\label{2-eq35}
\eea
Furthermore, we use the notation that ${\rm Extr}_{\Lambda}
\left\{g(\Lambda)\right\}
$ means the extremum with respect to $\Lambda$. Note that using the saddle-point method
to evaluate this by expanding the order parameters 
is comparatively tight for sufficiently large $N$.

From the extremum of the order parameters, we obtain
\bea
\label{eq41}
Q_{w}\eq\left\langle\left((\l\pmi) I_n+\tilde{Q}_w+ s \tilde{Q}_s\right)^{-1}\right\rangle_ s,\\
\label{1-eq25}
Q_s\eq\left\langle s 
\left((\l\pmi) I_n+\tilde{Q}_w+s\tilde{Q}_s\right)^{-1}\right\rangle_ s,\\
\label{1-eq22}\tilde{Q}_w\eq 0,\\
\label{1-eq23}
\tilde{Q}_s\eq\a(Q_s-I_n)^{-1}.
\eea
If we substitute \sref{1-eq22} and \sref{1-eq23} into 
\sref{eq41} and \sref{1-eq25}, then we have
\bea
Q_w\eq
\left\langle\left((\l\pmi)I_n+\a s  (Q_s-I_n)^{-1}\right)^{-1}\right\rangle_ s,\\
Q_s\eq
\left\langle s \left((\l\pmi)I_n+\a s (Q_s-I_n)^{-1}\right)^{-1}\right\rangle_ s.
\eea
From this solution, we assume the following replica-symmetric solution:
\bea
\label{eq47}
Q_w\eq \chi_wI_n+q_wD_n,\\
\label{1-eq93}
Q_s\eq \chi_sI_n+q_sD_n,
\eea
where $I_n$ is the identity matrix, and $D_n\in{\bf R}^{n\times n}$ is an $n$-dimensional 
square matrix in which each of the entries is unity.
From this, we obtain
\bea
\label{2-eq44}
\label{eq48}
\chi_w\eq\left\langle
\f{1}{\l\pmi+\f{\a s}{\chi_s-1} }
\right\rangle_ s,\\
\label{1-eq34}
q_w\eq\a
q_s
\left\langle
\f{1}{c}
\right\rangle_ s,\\
\label{2-eq46}
\label{eq50}
\chi_s\eq\left\langle
\f{s}{\l\pmi+\f{\a s}{\chi_s-1} }
\right\rangle_ s,\\
q_s\eq\a
q_s\left\langle
\f{ s}{c}
\right\rangle_ s,\label{1-eq35}
\eea
where $c=\f{1}{s}
((\l\pmi)(\chi_s-1)+\a s)
((\l\pmi)(\chi_s-1+nq_s)+\a s)$. Next, from \sref{1-eq34} and \sref{1-eq35}, we have
\bea
q_w\eq0,\\
q_s\eq0.
\eea
That is, the off-diagonal elements of 
$Q_w$ and $Q_s$ are estimated to be $0$. From \sref{2-eq44} and \sref{2-eq46}, 
since $\chi_w$ and $\chi_s$ do not depend on $n$, 
\sref{2-eq44} and \sref{2-eq46} hold for any $n$. Thus, by
using $\chi_s$ in \sref{eq50}, we can analytically assess
$\chi_w$ in \sref{eq48}.

Then, if we substitute $q_w=q_s=0$ into \sref{eq38-1}, we have
\bea
&&
\lim_{N\to\infty}\f{1}{N}
\log E_X
\left[Z^n(\l\pmi|X)\right]\nn
\eq\f{n\a}{2}\f{\chi_s}{\chi_s-1}
-\f{n\a}{2}\log(1-\chi_s)\nn
&&
-\f{n}{2}
\left\langle
\log
\left(
\l\pmi+\f{\a s}{\chi_s-1}
\right)
\right\rangle_s.
\eea
From this, we have the asymptotic limit
\bea
&&
\lim_{N\to\infty}\f{1}{N}
\log
E_X[Z^n(\l\pmi|X)]\nn
\eq
\lim_{N\to\infty}\f{n}{N}
\log
\left(E_X[Z(\l\pmi|X)]\right).
\eea
In a previous study \cite{Shinzato-SA2015}, 
it was found that this result implies that 
the distribution of the partition function is concentrated on a single point, 
the expectation of 
 the partition function. That is, roughly speaking, 
in the thermodynamic limit, for an arbitrary function of the partition function, 
$f(Z(\l\pmi|X))$, 
$E_X[f(Z(\l\pmi|X))]
=f(E_X[Z(\l\pmi|X)])
$ holds asymptotically. From this, we can assess 
the configurational average of the generating function, as follows:
\bea
E_X[\phi(\l\pmi|X)]\eq
\lim_{N\to\infty}\f{1}{N}\log E_X[Z(\l\pmi|X)]\nn
\eq
\f{\a}{2}\f{\chi_s}{\chi_s-1}
-\f{\a}{2}\log(1-\chi_s)\nn
&&
-\f{1}{2}
\left\langle
\log
\left(
\l\pmi+\f{\a s}{\chi_s-1}
\right)
\right\rangle_s.
\nn
\label{eq57}
\eea

Note that we do not let $n\to0$ in \sref{eq57}, but we take 
$\lim_{N\to\infty}\f{1}{N}E_X[\log Z(\l\pmi|X)]
=
\lim_{N\to\infty}\f{1}{N}
\log
E_X[Z(\l\pmi|X)]
$. From 
\sref{eq28} and 
\sref{eq57}, we obtain
$\rho(\l)
=-\f{1}{\pi}{\rm Im}
\lim_{\ve\to+0}
\chi_w$
for this case.

One point should be noted here. Based on a result obtained in {the previous work}, 
we should substitute \sref{eq36} and \sref{eq37} into \sref{eq10} 
and \sref{eq18} to obtain 
\bea
\chi_w\eq\f{1}{N}
\sum_{i=1}^N\f{1}{\l\pmi+\a\chi_t s_i},\\
\chi_s\eq\f{1}{N}
\sum_{i=1}^N\f{s_i}{\l\pmi+\a\chi_t s_i},\\
\chi_u\eq\f{1}{\chi_s-1},\\
\chi_t\eq\f{1}{\chi_s-1}.
\eea
In the limit of large $N$, the results obtained by our proposed replica 
approach are consistent with those obtained in {\cite{Burda1}}.

\subsection{Independent but not identically distributed; case 2}
We now consider the case in which the covariance is $E_X[x_{i\mu}x_{j\nu}]= t_\mu\d_{ij}\d_{\mu\nu}$, 
that is, we set
\bea
\label{eq61}
M\eq I_N\in{\bf R}^{N\times N},\\
\label{eq62}
\Theta\eq
{\rm diag}\left\{t_1,\cdots,t_p\right\}
\in{\bf R}^{p\times p},
\eea
and then proceed in a way similar to what we did in the previous subsection. That is, we begin by obtaining 
\bea
\label{2-eq57}
&&
\lim_{N\to\infty}
\f{1}{N}
\log E_X\left[Z^n(\l\pmi|X)\right]\nn
\eq
\mathop{\rm Extr}_{Q_w,\tilde{Q}_w,Q_t,\tilde{Q}_t}
\left\{
-\f{1}{2}
\log\det|
(\l\pmi)I_n
+\tilde{Q}_w
|
\right.
\nn
&&
+\f{1}{2}{\rm Tr}Q_w\tilde{Q}_w
-\f{\a}{2}{\rm Tr}Q_tQ_w
+\f{\a}{2}{\rm Tr}Q_t\tilde{Q}_t
\nn
&&
\left.
-\f{\a}{2}
\left\langle\log\det
|I_n- t \tilde{Q}_t|
\right\rangle_ t
\right\},
\eea
where  
\bea
\left\langle
f( t)
\right\rangle_ t\eq
\lim_{p\to\infty}\f{1}{p}\sum_{\mu=1}^p
f( t_\mu).\label{2-eq60}
\eea
From the extremum of \sref{2-eq57}, we obtain
\bea
\label{eq67}
Q_w\eq((\l\pmi)I_n+\tilde{Q}_w)^{-1},\\
\label{eq68}
\tilde{Q}_w\eq
\a Q_t,\\
\label{eq69}
Q_t\eq \left\langle
t\left( t \tilde{Q}_t-I_n\right)^{-1}
\right\rangle_ t,\\
\label{eq70}
\tilde{Q}_t\eq Q_w.
\eea
If we substitute 
\sref{eq69} and \sref{eq70} into \sref{eq67} and \sref{eq68}, we obtain 
simultaneous equations in terms of 
$Q_w$ and $\tilde{Q}_w$. 
Using $Q_w=\chi_wI_n+q_wD_n$ in \sref{eq47} and 
$\tilde{Q}_w=\tilde{\chi}_w I_n-\tilde{q}_wD_n$, we obtain the following saddle-point equations: 
\bea
\label{2-eq63}
\chi_w\eq\f{1}{\l\pmi+\tilde{\chi}_w},\\
\label{1-eq45}q_w\eq\f{\tilde{q}_w}{
(\l\pmi+\tilde{\chi}_w)
(\l\pmi+\tilde{\chi}_w-n\tilde{q}_w)
},\\
\label{2-eq65}\tilde{\chi}_w\eq
\a\left\langle
\f{t}{t\chi_w-1}
\right\rangle_ t,\\
\tilde{q}_w\eq\a
q_w
\left\langle
\f{t^2 }{( t\chi_w-1)
( t\chi_w-1+n t q_w)
}
\right\rangle_ t.
\label{1-eq47}
\eea
From \sref{1-eq45} and \sref{1-eq47}, we estimate 
$q_w=\tilde{q}_w=0$. Furthermore, since 
\sref{2-eq63} and \sref{2-eq65} hold for any $n$, 
$\lim_{N\to\infty}\f{1}{N}\log 
E_X[Z^n(\l\pmi|X)]=\lim_{N\to\infty}\f{n}{N}\log E_X[Z(\l\pmi|X)]$ 
holds approximately. From this, 
we can also find the asymptotic eigenvalue distribution for this case.

To compare this with the result of the previous {work \cite{Burda1}}, 
we
substitute \sref{eq61} and \sref{eq62} into \sref{eq21} to 
\sref{eq24} to obtain
\bea
\chi_w(=\chi_s)\eq\f{1}{\l\pmi+\a\chi_t},\\
\tilde{\chi}_w\eq\a\chi_t,\\
\chi_t\eq\f{1}{p}\sum_{\mu=1}^p
\f{t_\mu}{t_\mu\chi_s-1}.
\eea
That is, when $N$ is sufficiently large, the use of replica analysis in this case produces results that are consistent with those
obtained in previous studies.

\subsection{{Kronecker product correlation; case 3}}
As a more general case, 
we consider the asymptotic eigenvalue distribution 
when the covariance of the components of the random matrix is given by 
$E_X[x_{i\mu}x_{j\nu}]=m_{ij}\theta_{\mu \nu}$, that is, 
the components are mutually correlated. Since the 
correlation 
$E_X[x_{i\mu}x_{j\nu}]=m_{ij}\theta_{\mu \nu}$ is 
represented as a Kronecker product,
we can diagonalize 
$M=\left\{m_{ij}\right\}\in{\bf R}^{N\times N}$ and $\Theta=\left\{
\theta_{\mu\nu}
\right\}\in{\bf R}^{p\times p}$ with the diagonal matrices $S={\rm 
diag}\left\{ s_1,\cdots, s_N\right\}\in{\bf R}^{N\times N}$ and $T={\rm diag}
\left\{
 t_1,\cdots, t_p\right\}\in{\bf R}^{p\times p}$ and the orthogonal 
 matrices $W\in{\bf R}^{N\times N}
$ and $
U\in{\bf R}^{p\times p}$, such that $M=WSW^{\rm T}\in{\bf R}^{N\times N}$, 
$\Theta=U
TU^{\rm T}\in{\bf R}^{p\times p}$, and
\bea
&&\lim_{N\to\infty}\f{1}{N}
\log E_X[Z^n(\l\pmi|X)]
\nn
\eq
\mathop{\rm Extr}_{Q_w,Q_s,Q_u,Q_t,
\tilde{Q}_w,
\tilde{Q}_s,
\tilde{Q}_u,
\tilde{Q}_t
}
\left\{
-\f{\a}{2}{\rm Tr}Q_sQ_t
+\f{1}{2}{\rm Tr}Q_w\tilde{Q}_w
\right.\nn
&&
\left.
+\f{1}{2}{\rm Tr}Q_s\tilde{Q}_s+\f{\a}{2}{\rm Tr}Q_u\tilde{Q}_u
+\f{\a}{2}{\rm Tr}Q_t\tilde{Q}_t
\right.
\nn
&&
-\f{1}{2}
\left\langle
\log\det|
(\l\pmi)I_n+
\tilde{Q}_w+ s\tilde{Q}_s|
\right\rangle_{ s}
\nn
&&
\left.
-\f{\a}{2}
\left\langle
\log
\det
|I_n-\tilde{Q}_u
- t\tilde{Q}_t|
\right\rangle_{ t}
\right\}.
\label{1-eq54}
\eea
These are obtained using a similar approach to that used earlier in this paper (see Appendix \ref{app-a} for details). We
note that {this} is not dependent on either 
$W$ or $U$. From this, we obtain the saddle-point equations:
\bea
\label{eq78}
Q_w\eq
\left\langle
\left((\l\pmi)I_n+\tilde{Q}_w+ s\tilde{Q}_s\right)^{-1}
\right\rangle_ s,\\
Q_s\eq
\left\langle
 s\left((\l\pmi)I_n+\tilde{Q}_w+ s\tilde{Q}_s\right)^{-1}
\right\rangle_ s,\\
Q_u\eq
\left\langle
\left( t\tilde{Q}_t+\tilde{Q}_u-I_n\right)^{-1}
\right\rangle_ t,\\
\label{eq81}
Q_t\eq
\left\langle
 t\left( t\tilde{Q}_t+\tilde{Q}_u-I_n\right)^{-1}
\right\rangle_ t,\\
\tilde{Q}_w\eq0,\\
\tilde{Q}_s\eq\a Q_t,\\
\tilde{Q}_u\eq0,\\
\tilde{Q}_t\eq Q_s.
\eea
If we substitute 
$\tilde{Q}_w,\tilde{Q}_s,\tilde{Q}_u,\tilde{Q}_t$ into 
\sref{eq78} to \sref{eq81}, we obtain 
$Q_u=\chi_u I_n-q_uD_n$ and $Q_t=\chi_tI_n-q_tD_n$. In 
a similar way, since the off-diagonal elements of the order parameter matrices are $0$, we obtain
\bea
\label{2-eq76}
\chi_w\eq\left\langle
\f{1}{\l\pmi+\a s \chi_t}
\right\rangle_ s,\\
\label{2-eq77}
\chi_s\eq\left\langle
\f{ s}{\l\pmi+\a s \chi_t}
\right\rangle_ s,\\
\label{2-eq78}\chi_u\eq
\left\langle
\f{1}{ t\chi_s-1}
\right\rangle_ t,\\
\label{2-eq79}\chi_t\eq
\left\langle
\f{ t}{ t\chi_s-1}
\right\rangle_ t.
\eea
Note that 
this finding includes the findings presented in the previous subsection. Furthermore, 
we verified that the proposed method includes as a special case the approach based on Feynman diagrams. 
\section{Belief propagation algorithm\label{sec4}}
\subsection{Multivariate Gaussian distribution}

Replica analysis is one way to analyze a 
quenched ordered system by using self-averaging and/or the assumption 
that the matrix size $N$ is sufficiently large. However, 
an arbitrary 
random matrix ensemble is not always self-averaging, 
the size $N$ may be 
large but not infinite; for example, an assets 
return matrix in the mean-variance model of investment management is assumed to be finite, and so it is also important to be able to determine the eigenvalue distribution $\rho(\l|X)$ when $N$ is large but finite.

We use \sref{eq27}, as follows: 
\bea
\label{eq1-88}
&&
\rho(\l|X)\nn
\eq
\f{2}{\pi}{\rm Im}\lim_{\ve\to+0}
\pp{}{\l}\f{1}{N}
\log
\area 
\f{d\vec{w}
e^{-\f{1}{2}
\vec{w}^{\rm T}((\l\pmi)I_N-XX^{\rm T})
\vec{w}
}}{(2\pi)^{\f{N}{2}}}\nn
\eq-
\f{1}{\pi}
{\rm Im}\lim_{\ve\to+0}
\area d\vec{w}P(\vec{w}|\l,X)
\f{\vec{w}^{\rm T}\vec{w}}{N}.
\eea
The expectation of $\vec{w}^{\rm T}\vec{w}$ 
using $P(\vec{w}|\l,X)$ can be applied to determine the eigenvalue 
distribution, where the 
probability density function $P(\vec{w}|\l,X)$ is a
multivariate Gaussian distribution with $N$ variables:
\bea
\label{eq95}
P(\vec{w}|\l,X)\eq
\f{e^{-\f{1}{2}\vec{w}^{\rm T}
((\l\pmi)I_N
-XX^{\rm T})\vec{w}
}}{(2\pi)^{\f{N}{2}}
\det|(\l\pmi)I_N
-XX^{\rm T}|^{-\f{1}{2}}
}.\qquad
\eea

Note that since we must 
directly determine the inverse matrix and determinant of 
$(\l\pmi)I_N
-XX^{\rm T}$ in order to average $\vec{w}^{\rm T}\vec{w}$ using 
$P(\vec{w}|\l,X)$'Å$\vec{w}^{\rm T}\vec{w}$ in \sref{eq95}, 
when $N$ is large, the calculation time will be excessive. In order to 
reduce the required computation time, we will consider a way to assess the
expectation of $\vec{w}^{\rm T}\vec{w}$ with a trial distribution 
$Q(\vec{w})$ that, as evaluated by the Kullback-Leibler divergence, is approximately close to 
$P(\vec{w}|\l,X)$. 

\subsection{Derivation {from belief propagation} algorithm based on the Kullback-Leibler information criterion}
Based on the above discussion, 
we will derive $Q(\vec{w})$, which is an approximate trial distribution with respect to $P(\vec{w}|\l,X)$ and is
based on the Kullback-Leibler criterion \red{\cite{Kabashima}}.
In the context of belief propagation, 
$P(\vec{w}|\l,X)$ in \sref{eq95} is defined as follows:
\bea
P(\vec{w}|\l,X)
\eq
\f{1}{Z_P}\prod_{i=1}^NP_0(w_i)
\prod_{\mu=1}^p
g\left(
\f{\vec{x}_\mu^{\rm T}\vec{w}}{\sqrt{N}}
\right),\\
Z_P\eq
\area d\vec{w}
\prod_{i=1}^NP_0(w_i)
\prod_{\mu=1}^p
g\left(
\f{\vec{x}_\mu^{\rm T}\vec{w}}{\sqrt{N}}
\right)
,\qquad
\eea
where 
$P_0(w_i)=e^{-\f{\l\pmi}{2}w_i^2}$ and 
$g(v)=e^{\f{v^2}{2}}$. On the other hand, the 
trial distribution $Q(\vec{w})$ is defined using 
beliefs $b_i(w_i),b_\mu(\vec{w})$ as follows:
\bea
Q(\vec{w})
\eq
\f{1}{Z_Q}
\left(\prod_{i=1}^Nb_i(w_i)\right)^{1-p}
\prod_{\mu=1}^pb_{\mu}(\vec{w}),\\
Z_Q\eq\area d\vec{w}
\left(\prod_{i=1}^Nb_i(w_i)\right)^{1-p}
\prod_{\mu=1}^pb_{\mu}(\vec{w}),
\eea
where beliefs $b_i(w_i)$ and $b_\mu(\vec{w})$ are defined as
\bea
\forall i,\mu,\qquad
\label{eq100}
b_i(w_i)\eq\area \prod_{k=1,(k\ne i)}^Ndw_kb_\mu(\vec{w}),
\eea
where $\area\prod_{k=1,(k\ne i)}^Ndw_k$ means 
the integral with respect to $\vec{w}$ except for $w_i$. Thus, the
Bethe free energy, that is, the primary part of the Kullback-Leibler divergence between $P(\vec{w}|\l,X)$ and 
$Q(\vec{w})$ is
\bea
F\eq
\sum_{\mu=1}^p
\area d\vec{w}
b_\mu(\vec{w})\log
\left[
\f{b_\mu(\vec{w})}{
g\left(
\f{\vec{x}_\mu^{\rm T}\vec{w}}{\sqrt{N}}
\right)
\prod_{i=1}^NP_0(w_i)
}
\right]\nn
&&
+(1-p)
\sum_{i=1}^N\area dw_i
b_i(w_i)\log
\left[\f{b_i(w_i)}{P_0(w_i)}\right].
\eea
That is, we determine $b_i(w_i)$ and 
$b_\mu(\vec{w})$ such that they minimize the Bethe free energy under the constraint 
given by \sref{eq100}. Although we can derive more approximate trial 
distributions $Q(\vec{w})$, 
we would rather evaluate the mean and the variance of $w_i$ with $Q(\vec{w})$
instead of $Q(\vec{w})$ so that we can analytically assess the eigenvalue distribution.
From this, we obtain the mean and variance as follows:
\bea
\label{eq96}
m_{wi}\eq \area d\vec{w}Q(\vec{w})w_i,\\
\label{eq97}
\chi_{wi}\eq\area d\vec{w}Q(\vec{w})w_i^2-m_{wi}^2.
\eea

In this setting, we used a previously developed algorithm based on the belief propagation method 
\cite{Shinzato-Yasuda-BP2015,Roger}, and obtained the following:
\bea
\label{eq29}m_{wk}\eq\f{h_{wk}}{\l\pmi+\tilde{\chi}_{wk}},\\
\label{eq31}h_{wk}\eq\f{1}{\sqrt{N}}\sum_{\mu=1}^px_{k\mu}m_{u\mu}+\tilde{\chi}_{wk}m_{wk},\\
\label{eq98}m_{u\mu}\eq
\f{h_{u\mu}}{1-\tilde{\chi}_{u\mu}},\\
\label{eq35}h_{u\mu}\eq\f{1}{\sqrt{N}}\sum_{k=1}^Nx_{k\mu}m_{wk}-\tilde{\chi}_{u\mu}m_{u\mu},\\
\label{eq104}\chi_{wk}\eq\f{1}{\l\pmi+\tilde{\chi}_{wk}},\\
\label{eq105}\tilde{\chi}_{wk}\eq\f{1}{N}\sum_{\mu=1}^px_{k\mu}^2\chi_{u\mu},\\
\label{eq34}\chi_{u\mu}\eq\f{1}{\tilde{\chi}_{u\mu}-1},\\
\label{eq107}\tilde{\chi}_{u\mu}\eq\f{1}{N}\sum_{k=1}^Nx_{k\mu}^2\chi_{wk},
\eea
{
where we} note that the parameters other than $m_{wk}$ and $\chi_{wk}$ are 
auxiliary. It is easy to 
verify $m_{wk}=h_{wk}=m_{u\mu}=h_{u\mu}=0$, and 
we determine $\chi_{wk},\tilde{\chi}_{wk},\chi_{u\mu},\tilde{\chi}_{u\mu}$ such that they
satisfy \sref{eq104} to \sref{eq107}.

From \sref{eq1-88},
\sref{eq96}, and 
\sref{eq97}, the eigenvalue distribution $\rho(\l|X)$ is found to be
\bea
\rho(\l|X)\eq-\f{1}{\pi}{\rm Im}
\lim_{\ve\to+0}
\f{1}{N}
\sum_{i=1}^N
\chi_{wi}.
\eea
The complexity of this algorithm is estimated to be
$O(N^2)$, since the complexity of calculating the inverse matrix is $O(N^3)$, 
and thus our proposed approach is faster than the standard 
approach. 
This finding is consistent with 
an algorithm derived with the cavity method, in which the Bethe tree is assumed as the graphical model \cite{Roger}.

Finally, although we have considered a quenched 
disordered system, we also need to compare the results of our proposed method with 
those obtained in previous studies \red{\cite{Sengupta,Burda1}}. Thus, 
we rewrite \sref{eq104} as 
$\chi_{u\mu}=-1+\chi_{u\mu}\tilde{\chi}_{u\mu}$ and 
\sref{eq34} as $\chi_{wk}=\f{1-\chi_{wk}\tilde{\chi}_{wk}}{\l\pmi}$. 
It is then simple to evaluate the configurational average with respect to randomness in \sref{eq105} and 
\sref{eq107} by using $E_X[x_{k\mu}^2]=m_{kk}\theta_{\mu\mu}$, and we obtain 
\bea
\tilde{\chi}_{u\mu}\eq\theta_{\mu\mu}\chi_s,\\
\tilde{\chi}_{wk}\eq\a m_{kk}\chi_t,
\eea
where 
\bea
\label{eq112}
\chi_s\eq\f{1}{N}\sum_{k=1}^Nm_{kk}\chi_{wk},\\
\label{eq113}
\chi_t\eq \f{1}{p}\sum_{\mu=1}^p\theta_{\mu\mu}\chi_{u\mu}.
\eea
From this, we obtain
\bea
\label{eq114}
\chi_w\eq\f{1}{N}\sum_{k=1}^N\chi_{wk}\nn
\eq\f{1-\a\chi_t\chi_s}{\l\pmi},\\
\label{eq115}
\chi_u\eq\f{1}{p}\sum_{\mu=1}^p\chi_{u\mu}\nn
\eq-1+\chi_s\chi_t,
\eea
which correspond to \sref{eq20} and \sref{eq22}, respectively. In addition, 
$Q_w={\rm diag}\{\chi_{wk}\}\in{\bf C}^{N\times N}$ and 
$Q_u={\rm diag}\{\chi_{u\mu}\}\in{\bf C}^{p\times p}$, and from \sref{eq112} to \sref{eq115}, we obtain 
\bea
\chi_{w}\eq\f{1}{N}{\rm Tr}Q_w,\\
\chi_{u}\eq\f{1}{p}{\rm Tr}Q_u,\\
\chi_{s}\eq\f{1}{N}{\rm Tr}MQ_w,\\
\chi_{t}\eq\f{1}{p}{\rm Tr}\Theta Q_u;
\eea
these results are consistent with those obtained in previous studies \red{\cite{Sengupta,Burda1}}.
Note that if we are given
$E_X[x_{i\mu}x_{j\nu}]=m_{ij}\theta_{\mu\nu}$,
we can determine the asymptotic eigenvalue distribution 
with replica analysis and Feynman diagrams, and
when covariance is unknown, 
 we can determine the
eigenvalue distribution 
with the belief propagation algorithm, \sref{eq104} to 
\sref{eq107}. Note that this latter approach does not require knowledge of $E_X[x_{i\mu}x_{j\nu}]$.

\section{Numerical experiments and applications\label{sec5}}

We now consider the eigenvalue distributions obtained by the replica analysis and belief propagation algorithms,  
and we verify the proposed approaches by presenting the results of several numerical experiments.
\subsection{\red{Independent but not identically distributed; case 1}}
From the above arguments, 
since the mathematical structure of the second-order statistics of randomness is similar for all  
three cases
(for example, we can simultaneously diagonalize $M$ and $\Theta$ 
in replica analysis), we will \red{first} consider the 
independently but not identically distributed situation (case 1) in 
detail. 
We assume that the probability of $s_k$ follows the uniform distribution:
\bea
P(s_k)\eq\left\{
\begin{array}{ll}
\f{1}{s_{\max}-s_{\min}}&s_{\min}\le s_k\le s_{\max}\\
0&\text{otherwise}
\end{array}
\right.\label{eq88}
,
\eea
and we will consider the following three cases: case \red{(1,a)}: $(s_{\min},s_{\max})=(1,5)$;
case \red{(1,b)}: $(s_{\min},s_{\max})=(2,4)$; and case \red{(1,c)}: 
$(s_{\min},s_{\max})=(2.5,3.5)$ and $\a=p/N=4$.

\begin{figure}[h]
\begin{center}
\includegraphics[width=0.9\hsize,angle=0]{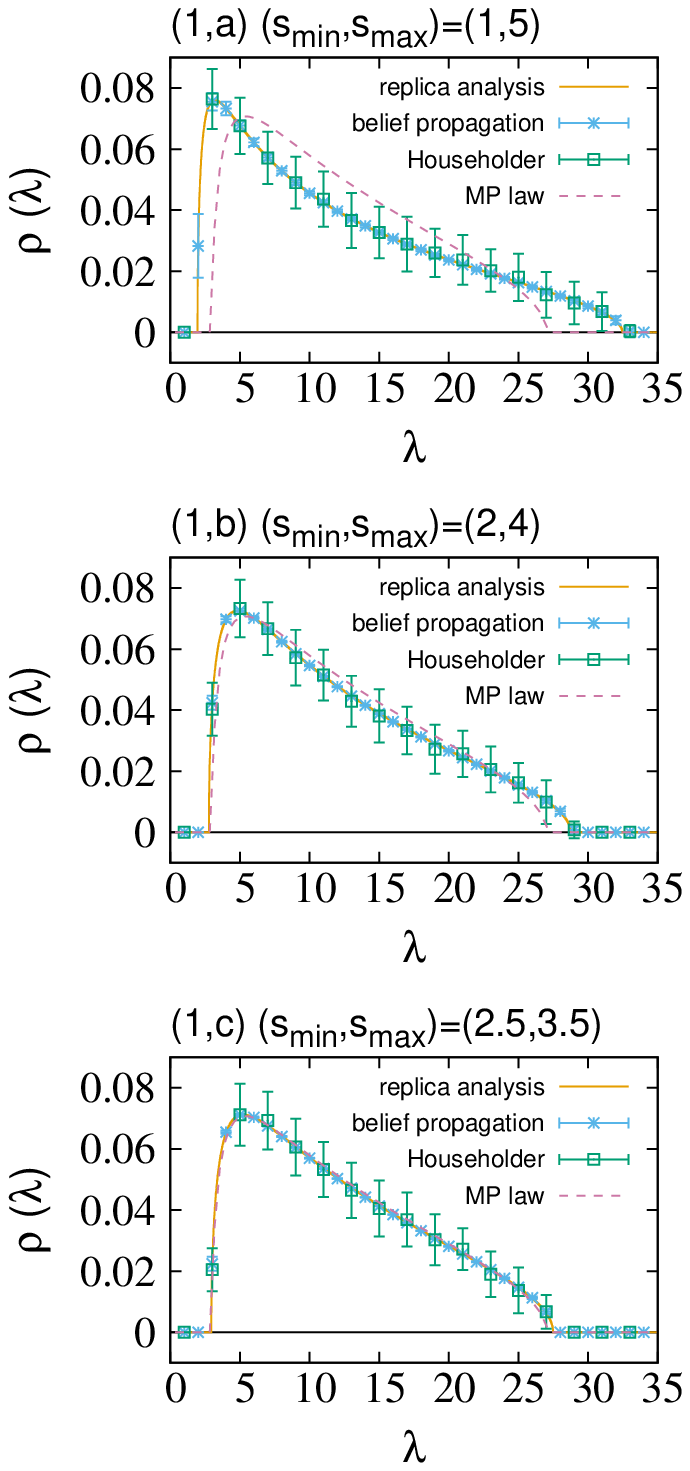}
\caption{
\label{Fig1}
\red{
Comparison of the asymptotic eigenvalue distribution derived with 
 analysis and by belief propagation for cases (1,a), (1,b) and (1,c), with $\a=p/N=4$.
The horizontal axis shows the eigenvalues, $\l$, and the vertical axis shows 
the asymptotic eigenvalue distribution, $\rho(\l)$.
The solid line (orange) shows the results of replica analysis, asterisks with 
 error bars (blue) show the results of belief propagation, and boxes with 
 error bars (green) show the results of the Householder method; the matrix 
 size is $N=500$ with $100$ samples. 
(1,a) $\l_{\min}\simeq1.950$ and $\l_{\max}\simeq32.487$. 
(1,b) $\l_{\min}\simeq2.768$ and $\l_{\max}\simeq28.762$. 
(1,c) $\l_{\min}\simeq2.944$ and $\l_{\max}\simeq27.504$.  As compared 
 with i.i.d. case, 
the dashed line (purple) shows the results of Mar$\check{\rm 
 c}$enko-Pastur law in \sref{eqMP}  with $v=\f{s_{\min}+s_{\max}}{2}=3$.
}
}
\end{center}
\end{figure}

The results are shown in 
Fig. \ref{Fig1}.
In order to verify the effectiveness of our proposed approaches, 
we compared the results with the
eigenvalue distributions derived from replica analysis and \red{belief 
propagation (see appendix \ref{app-b}).}
The matrix size used for the belief propagation experiments was $N=500$, and 
each component in the random matrix was assigned from the Gaussian distribution 
defined by hyperparameter $s_k$, which follows the random uniform 
distribution in \sref{eq88}; $100$ samples were prepared. 
As shown in \red{Fig. \ref{Fig1},} the results were in compliance with each other. 
In a similar manner, we used the Householder method (and the Sturm 
theorem) \red{\cite{Press}}, which 
can rigorously evaluate eigenvalue distributions; 
these results are also shown in \red{Fig. \ref{Fig1}.} 
For the Householder method, we plotted the average of $100$ samples with $N=500$. 
The results shown in \red{Fig. \ref{Fig1}} verify that the eigenvalue distributions 
can be accurately obtained with replica analysis and belief propagation, since 
they are consistent with the results of the Householder method.
\red{As compared with independently and identically distributed case, 
Mar$\check{\rm c}$enko-Pastur (MP) law when the component $x_{i\mu}$
is independently and identically distributed is defined as follows;
\bea
\label{eqMP}
\rho(\l)\eq
[1-\a]^+\d(\l)+
\f{\sqrt{[\l_+-\l]^+[\l-\l_-]^+}}{2\pi\l v},\qquad
\eea
where $\l_\pm=(1\pm\sqrt{\a})^2v$ and the constant 
$v=\f{1}{Np}\sum_{i=1}^N\sum_{\mu=1}^pE_X[x_{i\mu}^2]=\left\langle s\right\rangle_s
\left\langle t\right\rangle_t$ are used. For instance, if $v=3$ and 
$\a=4$, then $\l_-=3$ and $\l_+=27$. 
Shown in Fig. 
\ref{Fig1}, it turns out that when $|s_{\max}-s_{\min}|$ is becoming 
small, the eigenvalue distribution is close to MP law. In addition, the 
results of the previous works which handled the market correlation and 
analyzed the eigenvalues (and eigensignals) of financial cross-correlation matrix in detail are supported by our 
proposed methods \cite{Kwapien1,Kwapien,Drozdz}.
}
\subsection{\red{Independent but not identically distributed; case 2}}
\red{
Next, we also discuss another situation of independent but not identically distributed (case 2). 
We assume that the probability of $t_\mu$ follows the uniform 
distribution:
\bea
\label{eq116}
P(t_\mu)
\eq
\left\{
\begin{array}{ll}
\f{1}{t_{\max}-t_{\min}} &t_{\min}\le t_{\mu}\le t_{\max}\\
0&\text{otherwise}
\end{array}
\right.,
\eea
and we will consider the following three cases: case \red{(2,a)}: $(t_{\min},t_{\max})=(1,5)$;
case \red{(2,b)}: $(t_{\min},t_{\max})=(2,4)$; and case \red{(2,c)}: 
$(t_{\min},t_{\max})=(2.5,3.5)$ and $\a=p/N=4$. 
}

\begin{figure}[b]
\begin{center}
\includegraphics[width=0.9\hsize,angle=0]{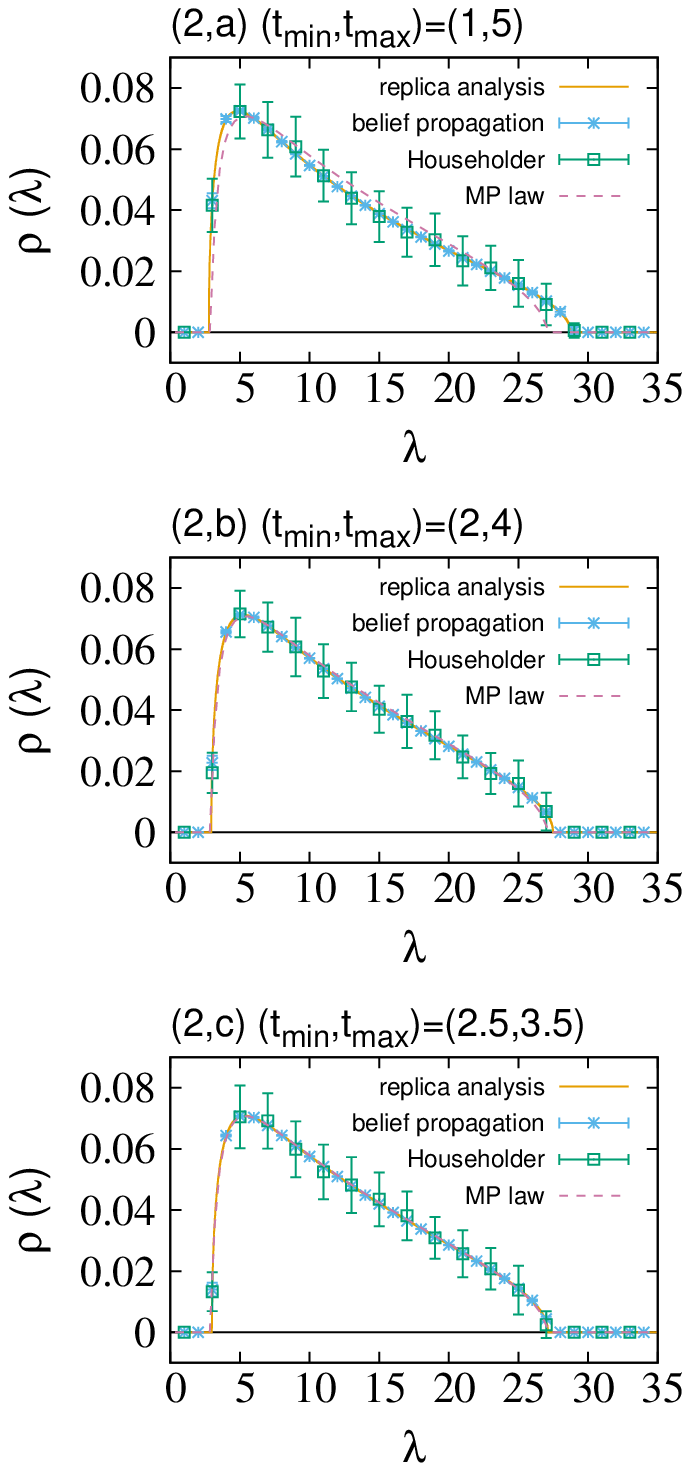}
\caption{
\label{Fig2}
\red{
Comparison of the asymptotic eigenvalue distribution derived with 
 analysis and by belief propagation for cases (2,a), (2,b), and (2,c), with $\a=p/N=4$.
The horizontal axis shows the eigenvalues, $\l$, and the vertical axis shows 
the asymptotic eigenvalue distribution, $\rho(\l)$.
The solid line (orange) shows the results of replica analysis, asterisks with 
 error bars (blue) show the results of belief propagation, and boxes with 
 error bars (green) show the results of the Householder method; the matrix 
 size is $N=500$ with $100$ samples. The numerical setting is similar to 
 that of Fig. \ref{Fig1}. 
(2,a) $\l_{\min}\simeq2.763$ and $\l_{\max}\simeq28.765$. 
(2,b) $\l_{\min}\simeq2.944$ and $\l_{\max}\simeq27.489$. 
(2,c) $\l_{\min}\simeq2.986$ and $\l_{\max}\simeq27.129$.  As compared 
 with i.i.d. case, 
the dashed line (purple) shows the results of Mar$\check{\rm 
 c}$enko-Pastur law in \sref{eqMP} with $v=\f{t_{\min}+t_{\max}}{2}=3$.
}
}
\end{center}
\end{figure}

\red{The results are shown in Fig. \ref{Fig2}. 
The effectiveness of our proposed approaches is verified from 
the comparation with the results of
the eigenvalue distributions from replica analysis and belief 
propagation (see appendix \ref{app-b}) and the one of Householder method. The numerical setting is similar to 
 that of case 1. As shown in Fig. \ref{Fig2}, the results were in 
 compliance with each other.  Moreover the eigenvalue distributions in 
 three cases are 
close to 
MP law in \sref{eqMP} with $v=\f{t_{\min}+t_{\max}}{2}=3$ because of the 
definition of Wishart matrix; its each element $(XX^{\rm T})_{ij}=\f{1}{N}\sum_{\mu=1}^px_{i\mu}x_{j\mu}$.
}

\subsection{\red{Kronecker product correlation; case 3}}
\begin{figure}[t]
\begin{center}
\includegraphics[width=0.9\hsize,angle=0]{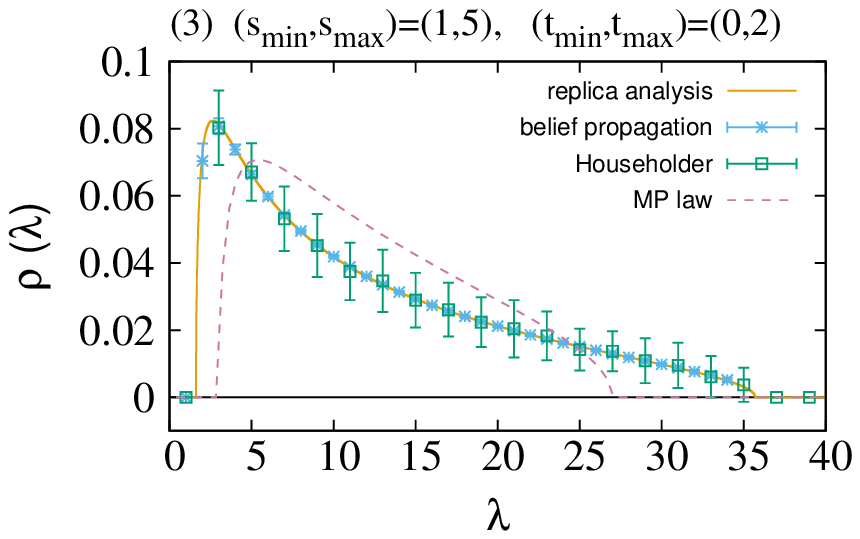}
\caption{
\red{
\label{Fig3}
Comparison of the asymptotic eigenvalue distributions derived by replica 
 analysis and belief propagation for case (3).
The numerical setting is similar to that of Fig. \ref{Fig1}.
(3) $\l_{\min}\simeq1.606$ and $\l_{\max}\simeq35.713$. 
As compared with i.i.d. case, 
the dashed line (purple) shows the results of Mar$\check{\rm 
 c}$enko-Pastur law in \sref{eqMP} with $v=\f{s_{\min}+s_{\max}}{2}\f{t_{\min}+t_{\max}}{2}=3$.
}
}
\end{center}
\end{figure}

\red{Lastly, we also discuss the situation of Kronecker product correlation 
(case 3). We use the parameter probabilities 
$P(s_k)$ in \sref{eq88} and 
$P(t_\mu)$ in \sref{eq116} with $(s_{\min},s_{\max})=(1,5)$ and 
$(t_{\min},t_{\max})=(0,2)$ because of 
$v=\f{s_{\min}+s_{\max}}{2}\f{t_{\min}+t_{\max}}{2}=3$. In 
Fig. \ref{Fig3}, it turns out that the results of three methods, replica analysis, belief propagation and 
Householder method,  are consistent. 
Futhermore, 
from case (1,a); $(s_{\min},s_{\max})=(1,5)$ and 
$(t_{\min},t_{\max})=(1,1)$ to case (3); $(s_{\min},s_{\max})=(1,5)$ and 
$(t_{\min},t_{\max})=(0,2)$, the smallest and largest eigenvalues are 
varied from $\l_{\min}(\simeq1.950)$ to $\l_{\min}(\simeq1.606)$ 
and from $\l_{\max}(\simeq32.487)$ to $\l_{\max}(\simeq35.713)$ as compared with $\l_-=3$ and $\l_+=27$ of MP law in \sref{eqMP}. 
}

\subsection{\red{Applications: Expectations of $\l^{-1}$ and $\l^{-2}$}}

Finally, we consider the expectations of $\l^{-1}$ and $\l^{-2}$ for 
this eigenvalue distribution in the independently but not identically 
distributed case, case 1. We begin with 
\bea
\left\langle
\f{1}{\l}
\right\rangle_\l
\eq\area d\l \rho(\l)\f{1}{\l}\nn
\eq-\f{1}{\pi}{\rm Im}\lim_{\ve\to+0}
\int_0^\infty dsP(s)
\area \f{d\l}{\l}\f{1}{\l\pmi+\f{\a s}{\chi_s(\l)-1}}\nn
\eq
-\f{1}{\pi}{\rm Im}\lim_{\ve\to+0}
\int_0^\infty dsP(s)\nn
&&
\lim_{R\to\infty,r\to+0}
\left(
\int_{-R}^{-r}{d\l}f(\l)
+\int_{r}^{R}{d\l}f(\l)
\right),
\eea
where, since $\chi_s$ is dependent on $\l$, we rewrite $\chi_s$ as 
$\chi_s(\l)$. Now, we have
\bea
f(\l)\eq
\f{1}{i\ve+\f{\a s}{\chi_s(\l)-1}}
\left[
\f{1}{\l}
-
\f{1}{\l\pmi+\f{\a s}{\chi_s(\l)-1}}
\right].\qquad
\eea
From the Cauchy integral theorem, we have
\bea
0\eq
\oint_C\f{dz}{z}\f{1}{i\ve+\f{\a s}{\chi_s(z)-1}}\nn
\eq
\lim_{
R\to\infty,r\to+0
}
\left[
\int_{-R}^{-r}\f{dz}{z}\f{1}{i\ve+\f{\a s}{\chi_s(z)-1}}
\right.\nn
&&
+\int_{r}^{R}\f{dz}{z}\f{1}{i\ve+\f{\a s}{\chi_s(z)-1}}
+i\int_\pi^0\f{d\theta}{i\ve+\f{\a s}{\chi_s(re^{i\theta})-1}}
\nn
&&
\left.
+i\int^\pi_0\f{d\theta}{i\ve+\f{\a s}{\chi_s(Re^{i\theta})-1}}
\right],\qquad
\eea
where we {already} replaced $z=re^{i\theta}$ in the third term 
and $z=Re^{i\theta}$ in the fourth term. Furthermore, since
\bea
\lim_{z\to0}\chi_s(z)\eq\f{1}{1-\a},\\
\lim_{|z|\to\infty}\chi_s(z)\eq1,
\eea
then
\bea
\lim_{\ve\to+0}
\area \f{dz}{z}\f{1}{i\ve+\f{\a s}{\chi_s(z)-1}}
\eq 
i\pi\f{\chi_s(0)-1}{\a s}\nn
\eq \f{i\pi}{s(1-\a)}.
\eea
Next, in a similar way, we estimate
\bea
\lim_{\ve\to+0}
\area \f{dz}{z
+{i\ve+\f{\a s}{\chi_s(z)-1}}
}\f{1}{i\ve+\f{\a s}{\chi_s(z)-1}}
\eq 0.\qquad
\eea
Thus, we obtain
\bea
\left\langle \f{1}{\l}
\right\rangle_\l
\eq\f{\left\langle s^{-1}
\right\rangle_s.
}{\a-1},\label{eq129}
\eea
Moreover, if we use
\bea
\lim_{z\to0}\pp{\chi_s(z)}{z}\eq
-\f{\a\left\langle
s^{-1}
\right\rangle_s
}{(\a-1)^3},\\
\lim_{|z|\to\infty}\pp{\chi_s(z)}{z}\eq0,
\eea
then we obtain
\bea
\left\langle \f{1}{\l^2}
\right\rangle_\l
\eq
\area d\l\rho(\l)\f{1}{\l^2}\nn
\eq
\f{\left\langle s^{-1}
\right\rangle_s^2
}{(\a-1)^3}+
\f{\left\langle s^{-2}
\right\rangle_s
}{(\a-1)^2}.
\label{eq132}
\eea
\red{
See appendix \ref{app-c}, they are consistent with the results from analysis of portfolio optimization problem.
}

\section{Summary and future work\label{sec6}}
In this paper, we considered the asymptotic eigenvalue distribution of a Wishart matrix defined by a random rectangular matrix. We considered three cases: 
{
(1) the components in each column are identically distributed, (2) the components in each row are not identically distributed, }
and (3) the components are correlated with one another. For each of these cases, we assessed the eigenvalue distribution using replica analysis, and we derived an algorithm for solving this based on belief propagation.
Our proposed approaches reproduced the findings of the 
Feynman diagram approach, which has been discussed in previous works, and the 
effectiveness of our approaches was validated by numerical experiments.

As an area of future work, since the random rectangular matrices considered in this paper can be regarded as dense, we also plan to analyze the asymptotic eigenvalue distribution for random rectangular sparse matrices, and to consider the case in which the entries are not identically distributed and that in which the entries are correlated with one another. In addition, since it is assumed in various applications (such as those in the cross-disciplinary fields of portfolio optimization, code division multiple access, and perceptron learning) that the components of the random rectangular matrix are i.i.d., our findings can be applied to the analysis of these problems, and the approaches discussed in previous works can be further developed for use with cases in which the entries are not i.i.d. and/or in which they are correlated.

\section*{Acknowledgements}
The author appreciates the fruitful comments of
 T. Nagao and K. Kobayashi. 
The work is in part supported by Grant-in-Aid Nos. 24710169 and 15K20999; the
President Project for Young Scientists at Akita Prefectural University; 
research project No. 50 of the National Institute of Informatics, Japan; 
research project No. 5 of the Japan Institute of Life Insurance; 
research project of the Institute of Economic Research Foundation; Kyoto University; 
research project No. 1414 of the Zengin Foundation for Studies in Economics 
and Finance; 
research project No. 2068 of the Institute of Statistical Mathematics; 
\red{research project of Mitsubishi UFJ Trust Scholarship Foundation; }
and research project No. 2 of the Kampo Foundation.

\appendix
\section{Replica calculation for the correlated case\label{app-a}}
When $E_X[x_{i\mu}x_{j\nu}]=m_{ij}\theta_{\mu\nu}$, we obtain
\bea
&&E_X[Z^n(\l\pmi|X)]\nn
\eq
\area 
\prod_{a=1}^n
\f{d\vec{w}_ad\vec{u}_ad\vec{v}_a
}{(2\pi)^{\f{Nn}{2}
+pn}}
E_X\left[
\exp
\left(
-\f{\l\pmi}{2}\sum_{a=1}^n\vec{w}_a^{\rm T}\vec{w}_a
\right.\right.
\nn
&&
\left.\left.
+\f{1}{2}
\sum_{a=1}^n
\vec{v}_a^{\rm T}
\vec{v}_a
+i\sum_{a=1}^n\vec{u}_a^{\rm T}\left(\vec{v}_a-X^{\rm T}\vec{w}_a\right)
\right)
\right]\nn
\eq
\area 
\prod_{a=1}^n
\f{d\vec{w}_ad\vec{u}_ad\vec{v}_a}{(2\pi)^{\f{Nn}{2}
+pn}}
\exp
\left(
-\f{\l\pmi}{2}\sum_{a=1}^n\vec{w}_a^{\rm T}\vec{w}_a
\right.\nn
&&
\left.
+
\sum_{a=1}^n
\left(\f{1}{2}
\vec{v}_a^{\rm T}
\vec{v}_a
+i
\vec{u}_a^{\rm T}
\vec{v}_a
\right)
-\f{p}{2}\sum_{a=1}^n
\sum_{b=1}^n
\f{\vec{w}_a^{\rm T}M\vec{w}_b}{N}
\f{\vec{u}_a^{\rm T}\Theta\vec{u}_b}{p}
\right).\nn
\eea
For the novel order parameters, we obtain 
{$q_{sab}=\f{\vec{w}_a^{\rm T}M\vec{w}_b}{N}$ and $
q_{tab}=\f{\vec{u}_a^{\rm T}\Theta\vec{u}_b}{p}$, }
then $\vec{z}_a= W^{\rm T}\vec{w}_a$ and $\vec{y}_a= U^{\rm T}\vec{u}_a$,
and we rewrite (A2) and (A3) as 
{$q_{sab}=\f{1}{N}\sum_{k=1}^Nz_{ia}z_{ib}s_i$ and $q_{tab}=\f{1}{p}\sum_{\mu=1}^py_{\mu a}y_{\mu b}t_\mu$,}
where $M=WSW^{\rm T}$ and $\Theta=UTU^{\rm T}$. From this, we obtain {$
q_{wab}=\f{1}{N}\sum_{i=1}^Nw_{ia}w_{ib}=\f{1}{N}\sum_{i=1}^Nz_{ia}z_{ib}
$ and $q_{uab}=\f{1}{p}\sum_{\mu=1}^pu_{\mu a}u_{\mu b}=\f{1}{p}\sum_{\mu=1}^py_{\mu a}y_{\mu b}$.}
Thus, we can evaluate 
\bea
&&
E_X[Z^n(\l\pmi|X)]\nn
\eq
\area \prod_{a=1}^n
\prod_{i=1}^N
\f{dw_{ia}
dz_{ia}
d\bar{z}_{ia}}{(2\pi)^{\f{3Nn}{2}}}
\prod_{a=1}^n
\prod_{\mu=1}^p
\f{
du_{\mu a}
dv_{\mu a}
dy_{\mu a}
d\bar{y}_{\mu a}
}{(2\pi)^{2pn}}
\nn
&&
\exp\left(
-\f{\l\pmi}{2}\sum_{i=1}^N\sum_{a=1}^nw_{ia}^2
\right.
+\f{1}{2}\sum_{\mu=1}^p\sum_{a=1}^nv_{\mu a}^2
\nn
&&
+i\sum_{\mu=1}^p\sum_{a=1}^nu_{\mu a}v_{\mu a}
-\f{p}{2}\sum_{a=1}^n\sum_{b=1}^n
q_{sab}q_{tab}
\nn
&&
+i\sum_{i=1}^N
\sum_{a=1}^n
\bar{z}_{ia}
\left(z_{ia}
-\sum_{k=1}^NW_{ik}^{\rm T}w_{ka}
\right)
\nn
&&
+i\sum_{\mu=1}^p
\sum_{a=1}^n
\bar{y}_{\mu a}
\left(y_{\mu a}
-\sum_{\nu=1}^pU_{\mu \nu}^{\rm T}u_{\nu a}
\right)\nn
&&-\f{1}{2}\sum_{a=1}^n\sum_{b=1}^n\tilde{q}_{wab}
\left(\sum_{i=1}^Nz_{ia}z_{ib}-Nq_{wab}
\right)
\nn
&&
-\f{1}{2}\sum_{a=1}^n\sum_{b=1}^n\tilde{q}_{sab}
\left(\sum_{i=1}^Nz_{ia}z_{ib}s_i-Nq_{sab}
\right)\nn
&&
-\f{1}{2}\sum_{a=1}^n\sum_{b=1}^n\tilde{q}_{uab}
\left(\sum_{\mu=1}^py_{\mu a}y_{\mu b}-pq_{uab}
\right)
\nn
&&
\left.
-\f{1}{2}\sum_{a=1}^n\sum_{b=1}^n\tilde{q}_{tab}
\left(\sum_{\mu=1}^py_{\mu a}y_{\mu b}t_\mu-pq_{tab}
\right)
\right)
.
\eea
Finally, we obtain
\bea
&&\lim_{N\to\infty}\f{1}{N}\log E_X\left[
Z^n(\l\pmi|X)
\right]\nn
\eq-\f{\a}{2}{\rm Tr}Q_sQ_t
+\f{1}{2}{\rm Tr}Q_s\tilde{Q}_s
+\f{1}{2}{\rm Tr}Q_w\tilde{Q}_w
+\f{\a}{2}{\rm Tr}Q_u\tilde{Q}_u\nn
&&+\f{\a}{2}{\rm Tr}Q_t\tilde{Q}_t-\f{1}{2}
\left\langle
\log\det\left|
(\l\pmi)I_n+\tilde{Q}_w+s\tilde{Q}_s
\right|
\right\rangle_s\nn
&&
-\f{\a}{2}
\left\langle
\log\det\left|
I_n-\tilde{Q}_u-t\tilde{Q}_t
\right|
\right\rangle_t,\qquad
\eea
where ${\rm Extr}$ is abbreviated {here}.

\section{\red{Algorithms based on replica analysis and belief propagation\label{app-b}}}
\red{We summary the both algorithms for resolving the eigenvalue 
distribution $\rho(\l)$ in three cases and use them in order to derive 
the eigenvalue distribution in numerical experiments.}
\red{
\subsection{Algorithms based on replica analysis}
\paragraph{Algorithm  for Case (1)}
In order to assess $\rho(\l)$ when $E_X[x_{i\mu}x_{j\nu}]=s_i\d_{ij}\d_{\mu\nu}$, we use the following iteration;
\bea
\chi_s\eq\left\langle
\f{s}{\l\pmi+\f{\a s}{\chi_s-1}}
\right\rangle_s,
\eea
then,  
\bea
\chi_w\eq\left\langle
\f{1}{\l\pmi+\f{\a s}{\chi_s-1}}
\right\rangle_s,\\
\rho(\l)\eq-\f{1}{\pi}
{\rm Im}\lim_{\ve\to+0}\chi_w.
\eea
\paragraph{Algorithm for Case (2)}
In order to assess $\rho(\l)$ when $E_X[x_{i\mu}x_{j\nu}]=t_\mu\d_{ij}\d_{\mu\nu}$, we use the following iterations;
\bea
\chi_w\eq\f{1}{\l\pmi+\tilde{\chi_w}},\\
\tilde{\chi}_w\eq\a
\left\langle
\f{t}{t\chi_w-1}
\right\rangle_t,
\eea
then,  
\bea
\rho(\l)\eq-\f{1}{\pi}
{\rm Im}\lim_{\ve\to+0}\chi_w.
\eea
\paragraph{Algorithm for Case (3)}
In order to assess $\rho(\l)$ when 
 $E_X[x_{i\mu}x_{j\nu}]=m_{ij}\theta_{\mu\nu}$; 
 $M=\left\{m_{ij}\right\}=WSW^{\rm T}\in{\bf R}^{N\times N}$ is composed 
 by the 
 diagonal matrix $S={\rm 
 diag}\left\{s_1,\cdots,s_N\right\}\in{\bf R}^{N\times N}$ and the 
 orthogonal matrix $W\in{\bf R}^{N\times N}$ and 
 $\Theta=\left\{\theta_{\mu\nu}\right\}=UTU^{\rm T}\in{\bf R}^{p\times p}$ is composed 
 by the 
 diagonal matrix $T={\rm 
 diag}\left\{t_1,\cdots,t_p\right\}\in{\bf R}^{p\times p}$ and the 
 orthogonal matrix $U\in{\bf R}^{p\times p}$,  we use the following iterations;
\bea
\chi_s\eq
\left\langle
\f{s}{\l\pmi+\a s\chi_t}
\right\rangle_s,\\
\chi_t\eq
\left\langle
\f{t}{t\chi_s-1}
\right\rangle_t,
\eea
then,  
\bea
\chi_w\eq
\left\langle
\f{1}{\l\pmi+\a s\chi_t}
\right\rangle_s,\\
\rho(\l)\eq-\f{1}{\pi}
{\rm Im}\lim_{\ve\to+0}\chi_w.
\eea
}

\red{
\subsection{Algorithm  based on belief propagation}
\paragraph{Algorithm for three cases}
In order to assess $\rho(\l)$, we use the following iterations;
\bea
\chi_{wk}\eq\f{1}{\l\pmi+\tilde{\chi}_{wk}},\\
\tilde{\chi}_{wk}\eq\f{1}{N}\sum_{\mu=1}^px_{k\mu}^2\chi_{u\mu},\\
\chi_{u\mu}\eq\f{1}{\tilde{\chi}_{u\mu}-1},\\
\tilde{\chi}_{u\mu}\eq\f{1}{N}\sum_{k=1}^Nx_{k\mu}^2\chi_{wk},
\eea
then,  
\bea
\chi_w\eq\f{1}{N}\sum_{k=1}^N\chi_{wk},\\
\rho(\l)\eq-\f{1}{\pi}
{\rm Im}\lim_{\ve\to+0}\chi_w.
\eea
}

\section{Two quantities in the portfolio optimization problem\label{app-c}}
From \cite{Shinzato-SA2015,Shinzato-heteroMV2016}, 
two quantities in portfolio optimization problem are derived by
replica analysis, as follows:
\bea
\ve\eq
\left\{
\begin{array}{l}
\f{1}{2\left\langle \l^{-1}
\right\rangle_\l
}\\
\f{\a-1}{2\left\langle s^{-1}\right\rangle_s}
\end{array}
\right.
,\\
q_w\eq
\left\{
\begin{array}{l}
\f{\left\langle \l^{-2}
\right\rangle_\l}{\left\langle \l^{-1}
\right\rangle_\l^2
}\\
\f{\left\langle s^{-2}\right\rangle_s}{\left\langle s^{-1}\right\rangle_s^2}
+\f{1}{\a-1}
\end{array}
\right.
,
\eea
where { $\epsilon$ and  $q_w$ are from \cite{Shinzato-SA2015,Shinzato-heteroMV2016} and the notation}
\bea
\left\langle f(\l)
\right\rangle_\l
\eq
\area 
d\l\rho(\l)f(\l).
\eea
From these, we can then find
\bea
\left\langle
\l^{-1}
\right\rangle_\l
\eq\f{\left\langle s^{-1}
\right\rangle_s
}{\a-1},\\
\left\langle
\l^{-2}
\right\rangle_\l
\eq\f{\left\langle s^{-1}
\right\rangle_s^2
}{(\a-1)^3}+
\f{\left\langle s^{-2}
\right\rangle_s
}{(\a-1)^2}
,
\eea
which are consistent with the results of \sref{eq129} and \sref{eq132}.

\end{document}